\documentclass[fleqn,usenatbib]{mnras}

\usepackage{newtxtext,newtxmath}

\usepackage[T1]{fontenc}

\DeclareRobustCommand{\VAN}[3]{#2}
\let\VANthebibliography\thebibliography
\def\thebibliography{\DeclareRobustCommand{\VAN}[3]{##3}\VANthebibliography}

\usepackage{graphicx}	
\usepackage{amsmath}	
\usepackage{booktabs}
\usepackage{pifont}

\newcommand\msun{$M$\mbox{$_{\normalsize\odot}$}}

\newcommand\lsun{$L$\mbox{$_{\normalsize\odot}$}}

\newcommand\logl{$\log (L/L_\odot)$}

\usepackage{soul}

\title[M31-2014-DS1]{The fate of the failed supernova candidate M31-2014-DS1}

\author[E.R. Beasor et al.]{
Emma R. Beasor$^{1}$\thanks{E-mail: E.R.Beasor@ljmu.ac.uk},
Nathan Smith$^{2}$,
Jeniveve Pearson$^{2}$,
Bhagya Subrayan$^{2}$,
Edo Berger$^{3}$,
David J. Sand$^{2}$,\newauthor
and Jay Strader$^{4}$ \\
$^{1}$Astrophysics Research Institute, Liverpool John Moores University, IC2, 146 Brownlow Hill, Liverpool, L3 5RF, UK \\
$^{2}$Steward Observatory, University of Arizona, 933 North Cherry Avenue, Tucson, AZ 85721-0065, USA \\
$^{3}$Center for Astrophysics \textbar{} Harvard \& Smithsonian, 60 Garden Street, Cambridge, MA 02138, USA \\
$^{4}$Department of Physics and Astronomy, Michigan State University, East Lansing, MI 48824, USA
}

\date{Accepted XXX. Received YYY; in original form ZZZ}

\pubyear{\the\year{}}

\usepackage{tabularx}
\begin{document}
\label{firstpage}
\pagerange{\pageref{firstpage}--\pageref{lastpage}}
\maketitle

\begin{abstract}
The fate of massive stars above 20\msun\ remains uncertain. Debate persists about whether they die as supernovae (SNe), or if they collapse directly into black holes (BHs) with little or no optical outburst — so-called "failed supernovae". The source M31-2014-DS1 experienced an optical outburst in 2014 and has remained faint at visual wavelengths since then. Due to its persistent faintness, it has been proposed as a failed SN candidate. We present new observations of this candidate obtained using the James Webb Space Telescope (JWST), the Submillimeter Array (SMA), and Chandra.  The {\it JWST} observations demonstrate that a luminous mid-infrared source persists at the same location a decade after the star faded at visual wavelengths. We model its current spectral energy distribution (SED) as a dust-enshrouded star.  No X-ray emission is detected, disfavoring the hypothesis that the late-time luminosity is powered by accretion onto a BH. We find that the remaining source is highly obscured by an asymmetric distribution of circumstellar dust, making it difficult to quantify its physical properties using spherically symmetric radiative transfer codes. The dust geometry requires that the inferred bolometric luminosity is only a lower limit, as a significant fraction of the central source's radiation may escape without being reprocessed by dust. We discuss the implications of these findings in the context of failed SN models and consider the potential overlap with signatures expected from a stellar merger, which also seems to provide a plausible explanation of this source.
\end{abstract}




\section{Introduction}
In the canonical view of single-star evolution, solar metallicity stars with initial masses above 8\msun\ are expected to end their lives as supernovae (SNe) and leave behind a neutron star (NS) remnant \citep{heger2003how}. However, the existence of stellar-mass black holes (BHs) requires that some massive stars collapse further to make remnants more massive than NSs.  This possibility has inspired suggestions that some of these massive stars may fail to explode successfully, disappearing from view rather than producing a bright transient \citep{kochanek2008survey}. The factors determining how a massive star ends its life are complex, and likely not a simple function of initial mass \citep[e.g.][]{sukhbold2018high, laplace2025bhs}. 

Serendipitious pre-explosion imaging of the progenitors to supernovae (SNe) has played an important role in our understanding of how massive stars end their lives, as it is arguably the most direct method of linking SNe to their progenitors. It was through this direct detection technique that red supergiants (RSGs) were confirmed as the most common progenitors to Type II-P supernovae \citep[e.g.][]{smartt2004detection,maund2005hubble, fraser2016disappearance}. However, as the sample of Type II-P SNe with pre-explosion imaging increased, some authors noted an apparent lack of progenitors with inferred initial masses above 16-18 $M_{\odot}$, despite the existence of RSGs in nearby stellar populations that are thought to have initial masses in this range \citep{smartt2009death, smartt2015observational}. The statistical significance of the putatively missing high-mass progenitors is controversial, however, and has been widely debated \citep[][among others]{davies2018initial, davies2020red, davies2020on, kochanek2020on, beasor2025luminosity, healy2024rsg}. In addition to the low statistical significance of missing progenitors, there are a number of other caveats to the RSG problem. For one, the initial masses are derived using stellar evolution models, which are highly uncertain at late phases \citep[e.g.,][]{renzo2017systematic,zapartas2021effect,eldridge2022review}. Secondly, progenitor luminosities are typically estimated using only the HST F814W filter, which can underestimate the true luminosity \citep{beasor2025luminosity}. Third, in the original RSG problem analysis \citep{smartt2009death}, no correction was applied for circumstellar dust extinction \citep{walmswell2012circumstellar}, which can further bias progenitor luminosities downward. Finally, most studies have focused on Type II-P supernovae, but higher-mass RSGs may preferentially explode as other SN types, such as Type IIn or stripped-envelope SNe \citep[e.g.,][]{smith2011observed}, which could change the inferred progenitor mass distribution.

The lack of high-mass RSG progenitors has been interpreted by some  as a smoking gun suggesting that the most luminous RSGs collapse to form a BH with little or no explosion \cite[e.g.][]{smartt2009death, smartt2015observational}. Following this observational result, theoretical work identified possible regimes of stellar mass which seem to be harder to explode in simulations, leading to the suggestion that there are `islands of explodability' \citep[e.g.][]{sukhbold2018high,couch2020sne,ebinger2019explodability,laplace2025bhs}, in which a high mass star is less likely to successfully explode\footnote{We note that this result comes from 1-D stellar evolution modeling, and explodability is determined by adopting a criterion for the progenitors core compactness. The use of this simple compactness parameter to determine whether or not a progenitor will explode is debated, and the core structure may be more complicated than 1-D stellar evolution models predict.}. Some recent work has suggested that these studies overestimate the likelihood of direct collapse in the RSG mass regime \citep[e.g.][]{burrows2024sne, burrows2025sne}. There is also support from the gravitational wave side, with the mass range of detected BHs implying a lower initial-mass transition from neutron star (NS) to BH \citep[e.g.][]{schneider2025}, as well as some works claiming that there is no lower BH mass gap at all \citep{ray2025bhmassgap}.

All of this has motivated the search for the observational signature of a failed SN. Generally, teams search for failed SNe the same way they search for successful SNe; by searching for sudden changes in brightness. In the failed SN searches, however, a progenitor star must first be detected in multiple epochs before it can be seen to fade or disappear entirely. As such, the search is limited to very nearby galaxies \citep{kochanek2008survey}. 

There are two key challenges in identifying failed SNe.  First, massive stars might fade for reasons other than collapse to a BH, so there may be other types of fading stars that masquerade as failed SNe \citep[see][]{jencson2022exceptional,beasor2024BH1}.  And second, from various theoretical predictions, it is unclear exactly what a failed SN should look like. There is a wide range of theoretical predictions for their observable characteristics. For one, it has been suggested that  failed SNe may undergo a direct collapse and have no visible signature beyond disappearing \citep{kochanek2008survey}. \citet{lovegrove2013very} suggest the collapse of a massive star may be accompanied by lingering infrared emission, potentially lasting for several years, but would not expect an optical transient. \citet{perna2014disk} suggest a failed supernova would produce a dim but prolonged optical transient due to long-lasting fallback accretion onto the BH, while \citet{antoni2023failed} find the collapse of a massive star into a BH would produce a bright transient due to the luminosity from rapid accretion. The inherent ambiguity in the wide range of current theoretical predictions makes it challenging to uniquely identify a fading star as a failed SN, especially since other types of events like stellar mergers may look similar \citep[see discussion in][]{beasor2024BH1}. 
 
The most compelling candidate of a failed SN to date had been N6946-BH1 (hereafter BH1), first identified by \citet{adams2017search}. The progenitor to the source was initially interpreted as an RSG with an initial mass of $\sim$30\msun, which then underwent an optical outburst before fading by a factor of 10 over 3000 days. However, it has also been noted that a number of observations prior to the outburst were more similar to those expected for stellar mergers \citep[e.g.][]{beasor2024BH1}. For one, the source had appeared to redden over the preceeding 5 years from the colors of a warm supergiant to those of a cool, red supergiant \citep[e.g.][]{adams2017search, humphreys2019bh1,beasor2024BH1}\footnote{We note that the reddening timescale may be longer, but there are no Spitzer data before 2004.}, similar to known stellar merger event V1309 Sco \citep{shara2010v1309sco,mason2010v1309sco,ste2011v1309sco}. Further, some models suggest the optical outburst of BH1 may have been too faint to be from a failed SN \citep{antoni2023failed}. Finally, there remains an IR bright source at the position of BH1. As discussed in \citet{beasor2024BH1} and \citet{kochanek2024bh1}, it is not possible to distinguish between the failed SNe and merger scenarios without further epochs of data. 

There have been other proposed candidates for failed SNe, including M101-OC1, a blue supergiant, \citep{neustadt2021lbt}, N3021-OC1, a yellow supergiant \citep{reynolds2015gone} and PHL293B, a luminous blue variable \citep[LBV][]{allan2020phl}. We note that neither of these objects have been considered to be strong candidates for failed SNe and have not been followed up with mid-IR facilities to confirm their disappearance. There is emerging evidence that some massive stars undergo major dimming events before re-brightening at later epochs, including M51-DS1 \citep{jencson2022exceptional} and Betelgeuse \citep{montarges2021betelgeuse}. The cause of these dimming events is unclear, but these studies show that any failed SNe candidate needs to be followed up at later times to confirm the source has not re-appeared.  

Recently, \cite{de2024ds1} identified a new failed SN candidate in M31. This new candidate was identified by searching for outbursts in data from the NEOWISE mid-infrared survey \citep{neowise} in the direction of Andromeda (M31). De et al. identified a source that brightened by 50\% in the mid-IR over two years beginning in 2014. For the subsequent thousand days after the MIR brightening started, the inferred total luminosity remained nearly constant, before fading below the progenitor flux over the next year, and NEOWISE data showed it continued to fade through 2022. During this time the optical flux from the source faded by a factor of $\geq$ 100 between 2016 and 2019. The authors dubbed the candidate M31-2014-DS1, and claim its moderate outburst and subsequent fading can be explained by the infall of the stellar envelope onto a newly formed black hole. The progenitor star was interpreted as a hydrogen-poor supergiant, with an inferred luminosity of \logl\ = 5 and effective temperature of $\sim$ 4500K \citep{de2024ds1}. We note that there are a number of SN with detected progenitors at a similar luminosity \citep[see Fig. 6 in][and references within]{beasor2025luminosity}, implying that stars at this luminosity are able to yield successful explosions in many cases. Also, if the star is a stripped-envelope progenitor \citep[as modelled by][]{de2024ds1}, some theoretical work suggests these stars should be easier to explode than their non-stripped counterparts \citep[e.g.][]{laplace2025bhs}. 
Here, we present new $JWST$ and $SMA$ observations of M31-2014-BH1 and model the data using ${\tt DUSTY}$ modelling. In Section 2 we discuss the near-IR, mid-IR, radio and X-ray data. In Section 3 we discuss the dust shell model fitting. In Section 4 we discuss current failed SN models. In Sections 5 \& 6 we discuss whether the observations of M31-2014-DS1 are consistent with either the failed SNe model or stellar mergers.

\section{Data}\label{sec:data}

\subsection{JWST}

The position of M31-2014-DS1 was observed by the James Webb Space Telescope (JWST) on November 28 2024. The data included NIRSpec \citep{nirspec} and MIRI LRS \citep{miri2015} spectra as well as photometry from and MIRI, taken under DDT Program 6809 \citep[PI: De,][]{deDDT}. 

\begin{figure*}
        \centering
        \includegraphics[width=\linewidth]{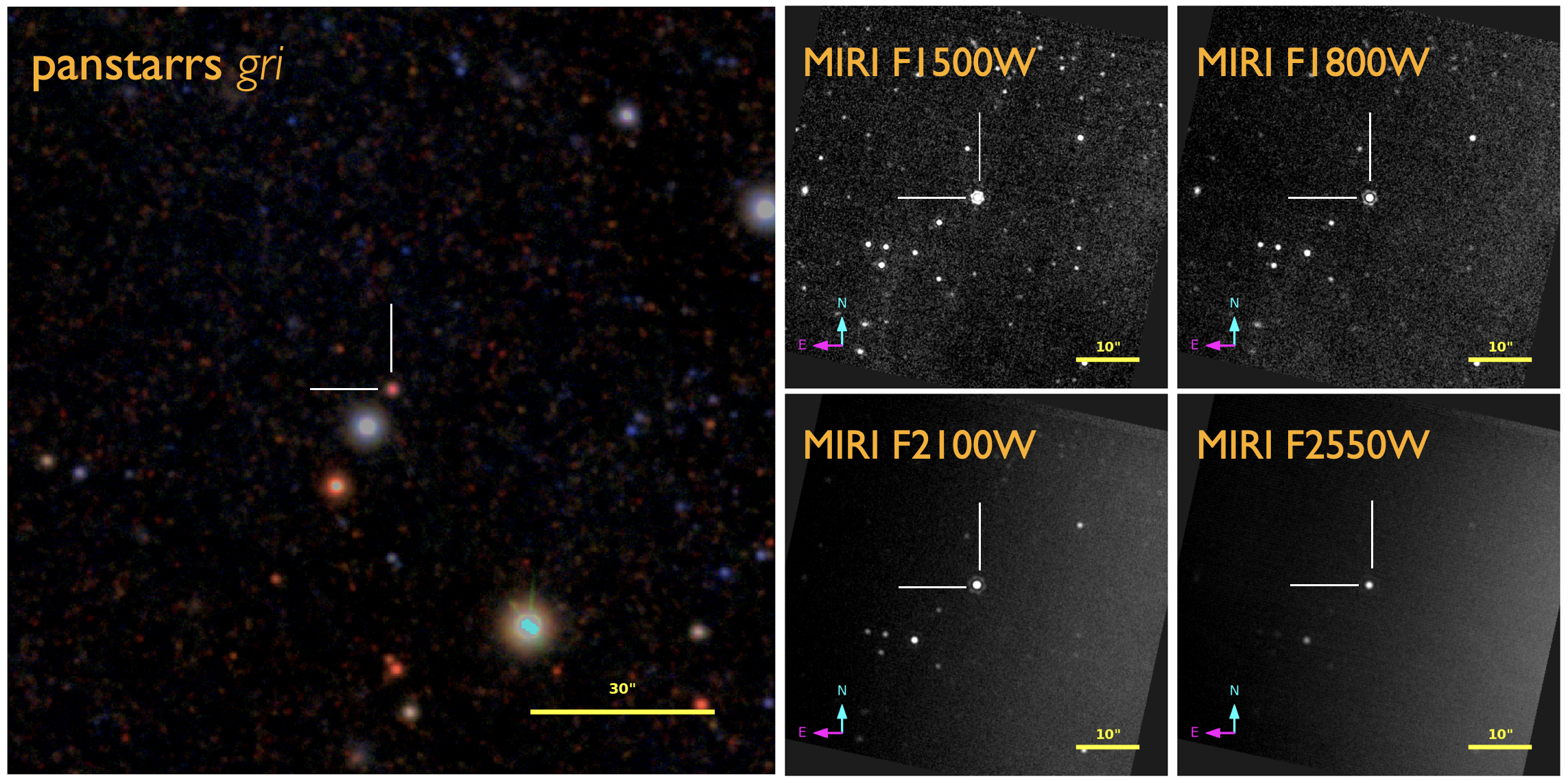}
    \includegraphics[width=\linewidth]{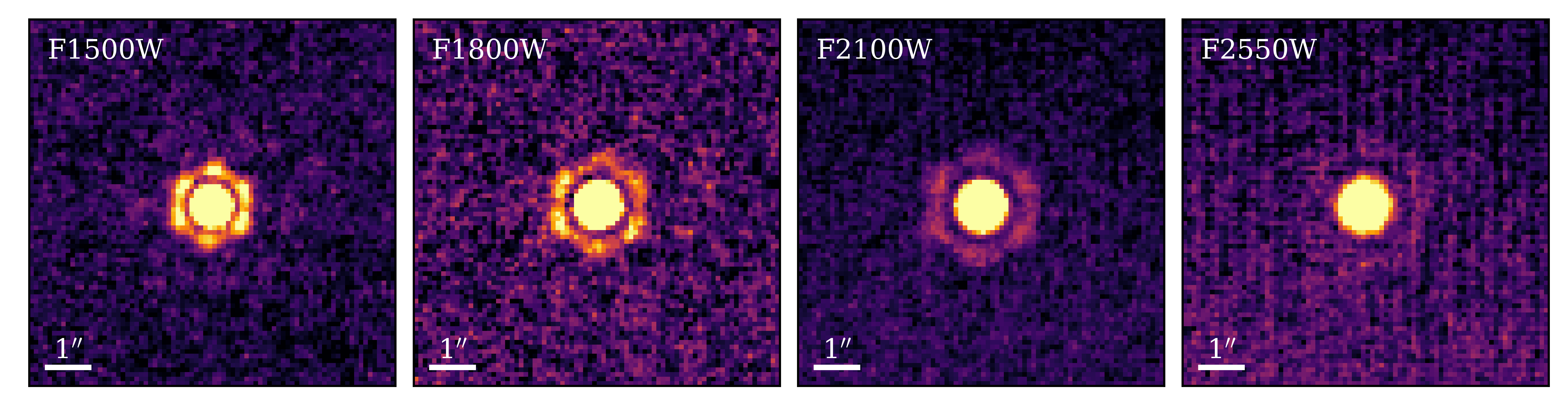}
    \caption{{\it  Top left:} Finder chart for M31-2014-DS1 using Pan-starrs \textit{gri} photometry. {\it Top Right:} JWST/MIRI imaging of the source remaining at the position of M31-2014-DS1. {\it Bottom:} PSFs in each JWST MIRI filter. } 
    \label{fig:miricutouts}
\end{figure*}

Imaging was taken in MIRI filters F1500W, F1800W, F2100W and F2500W, see Fig. \ref{fig:miricutouts}. We identified the source using the bluest MIRI filter and registered all images to a pre-disappearance HST image of the field, which served as the reference position. Alignment was done manually by eye, allowing us to confirm that the same source was clearly detected in all subsequent observations. Following the method described in \citet{pearson2025eaw}, PSF photometry of the MIRI observations was done using \texttt{space\textunderscore phot}\footnote{\texttt{space\textunderscore phot} version 0.2.5 \url{space-phot.readthedocs.io}} \citep{spacephot, Pierel24} on the Level 2 products. This is done by fitting the supernova's PSF in each of the Level 2 CAL files using WebbPSF \citep[version 1.2.1]{WebbPSF1, WebbPSF2} models. Given the low signal to noise detection of the source in the F2550W individual exposures, for this filter we instead do PSF photometry on the Level 3 stacked images. For photometry on the Level 3 products, \texttt{space\textunderscore phot} uses temporally and spatially dependent Level 2 PSF models from WebbPSF, and drizzles them together to create a Level 3 PSF model. The resulting MIRI photometry is listed in Table \ref{tab:miri} and MIRI images are shown in Figure \ref{fig:miricutouts}.

The NIRSpec data were obtained via the MAST data archive \citep{MAST} and were processed using the standard data reduction pipeline. Data were taking in the fixed slit mode using the G140H/F100LP, G235H/F170LP and G395H/F290LP gratings resulting in spectral coverage bwteeen $1.0\,\mu$m and $5.0\,\mu$m. The 1-D spectra were  extracted using the {\tt spec1d} function from {\tt specutils} \citep{specutils}. We then smoothed the spectrum using a moving average and a window size of 5 pixels. 

The JWST/MIRI LRS data were retrieved from the MAST archive and processed using the standard JWST pipeline which utilizes the \texttt{calwebb\_detector1} and \texttt{calwebb\_spec2} stages to correct detector artifacts and perform calibration \citep{bushouse_2022_7229890}. The spectral range covered is 5 to 14$\mu$m. One-dimensional spectra are extracted using \texttt{extract\_1d} summing the flux within a defined spatial aperture along the source trace, optimizing flux recovery. Background subtraction was carried out by estimating the background flux from regions adjacent to the source trace and subtracting it from the flux measured within the extraction aperture.\footnote{\url{https://github.com/STScI-MIRI/LRS_ExampleNB/blob/main/miri_lrs_pipeline_extraction.ipynb}}

\begin{figure*}
    \centering
    \includegraphics[width=\linewidth]{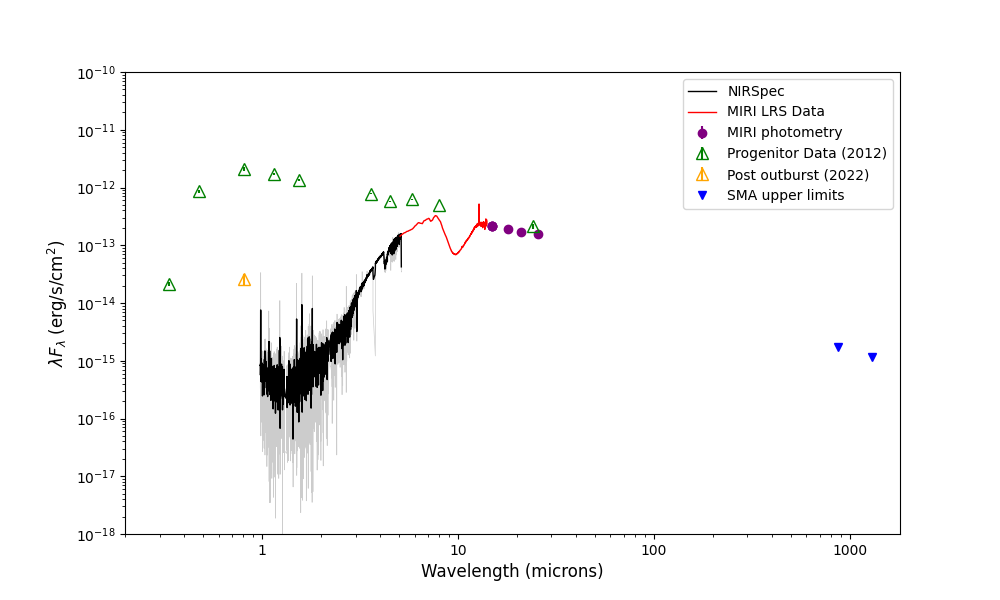}
    \caption{The SED of the remaining source at the position of M31-2014-DS1 obserevd by JWST and SMA. The NIRSpec data have been smoothed using a moving average with a window size of 5 pixels (black solid line). We also show the SED of the progenitor star (green triangles) as well as post-outburst data (orange triangle).}
    \label{fig:progenitor}
\end{figure*}
\begin{table}
\centering
\begin{tabular}{cc}
\hline
\textbf{Filter} & \textbf{Flux ($\times$ 10$^{-6}$ mJy)} \\
\hline
F1500W & 1.0902 $\pm$ 0.0059\\
F1800W & 1.1295 $\pm$ 0.0136  \\
F2100W & 1.1718 $\pm$ 0.0148  \\
F2550W & 1.3342 $\pm$ 0.0137  \\
\hline
\end{tabular}
\caption{Flux measurments for MIRI photometry.}
\label{tab:miri}
\end{table}
\subsection{Radio}
The Submillimeter Array (SMA) observations were obtained as part of the Large Project POETS (Pursuit of Extragalactic Transients with the SMA; project 2022B-S046, PI: Berger). The observations are summarized in Table \ref{tab:sma}. During these observations, the SMA was tuned to a local oscillator (LO) frequency of 225.5 GHz, providing spectral coverage at $209.5–221.5$ and $229.5–241.5$ GHz. Across all nights, 3C84 was observed as a bandpass calibrator, Uranus was observed as a flux calibrator, and J0136+478 was observed as a gain calibrator, with a 12-min cycle time cadence.

Analysis of the data was performed using the SMA COMPASS pipeline (G. K. Keating et al. 2025, in prep.), which flags spectral data based on outliers in amplitude when coherently averaging over increasing time intervals for each channel within each baseline, as well as baselines where little to no coherence is seen on calibrator targets. Flux calibration was performed using the Butler–JPL–Horizons 2012 (Butler 2012) model for Uranus. The data were imaged, and deconvolution was performed via the CLEAN algorithm (Högbom 1974).

We do not detect emission at the location of M31-DS1 in any of our observations, with a combined rms noise level of 0.5 mJy. 

\begin{table}
\centering
\begin{tabular}{cc}
\hline
\textbf{Observation Date} & \textbf{On-source Time (minutes)} \\
\hline
250219\_04:11:12 & 79 \\
250218\_03:56:04 & 130 \\
250217\_03:46:50 & 139 \\
250216\_04:05:22 & 79 \\
\hline
\end{tabular}
\caption{SMA observations.}
\label{tab:sma}
\end{table}
\subsection{X-ray}
X-ray observations at the position of M31-2014-DS1 were taken with ACIS-S \citep{Garmire2000} on board the {\it Chandra} X-ray observatory under DDT program 25509003 (PI: De) on 6 November 2024. The data were reprocessed with \texttt{CIAO} v4.17 and CALDB 4.12.2. Source counts were extracted from a $2''$ circular aperture centered on the optical position of M31-2014-DS1, and the background was estimated from a concentric annulus with inner and outer radii of $5''$ and $10''$. Spectra and instrumental responses were generated using \texttt{specextract}. The on source time was 10734.8 s and no counts were detected in either the source or background regions, leading to a 90\% confidence upper limit of $<2.1\times10^{-4}$ counts s$^{-1}$.

To convert the observed count rates into fluxes, we assumed an absorbed power-law spectrum with $N_\mathrm{H} = 1\times10^{23}\,\mathrm{cm^{-2}}$ and photon index $\Gamma = 1.7$. The adopted value of $N_\mathrm{H}$ reflects the expected high intrinsic extinction toward the source \citep{bahramian2015nh}. Under these assumptions, we derive an upper limit on the X-ray luminosity of $L_X < 2.96\times10^{35}\,\mathrm{erg\,s^{-1}}$ in the 0.5--7.0 keV band.

\section{The Spectral Energy Distribution}
The JWST spectral energy distribution (SED) of M31-2014-DS1 is shown in Fig. \ref{fig:progenitor}. The NIRSpec data reveal that at short optical/near-IR wavelengths, a faint, red source persists at the position of the failed SN candidate whereas MIRI data reveal a luminous mid-IR source that peaks around 8-10 $\micron$, which has almost the same 20 $\mu$m flux as the progenitor. At the shortest wavelengths, there is also a slight uptick in the flux toward shorter wavelengths ($\sim$ 1-2$mu$m) extending past the observed wavelength range,  perhaps suggesting the presence of a lingering optical source.

\begin{figure}
        \centering
    \includegraphics[width=\columnwidth]{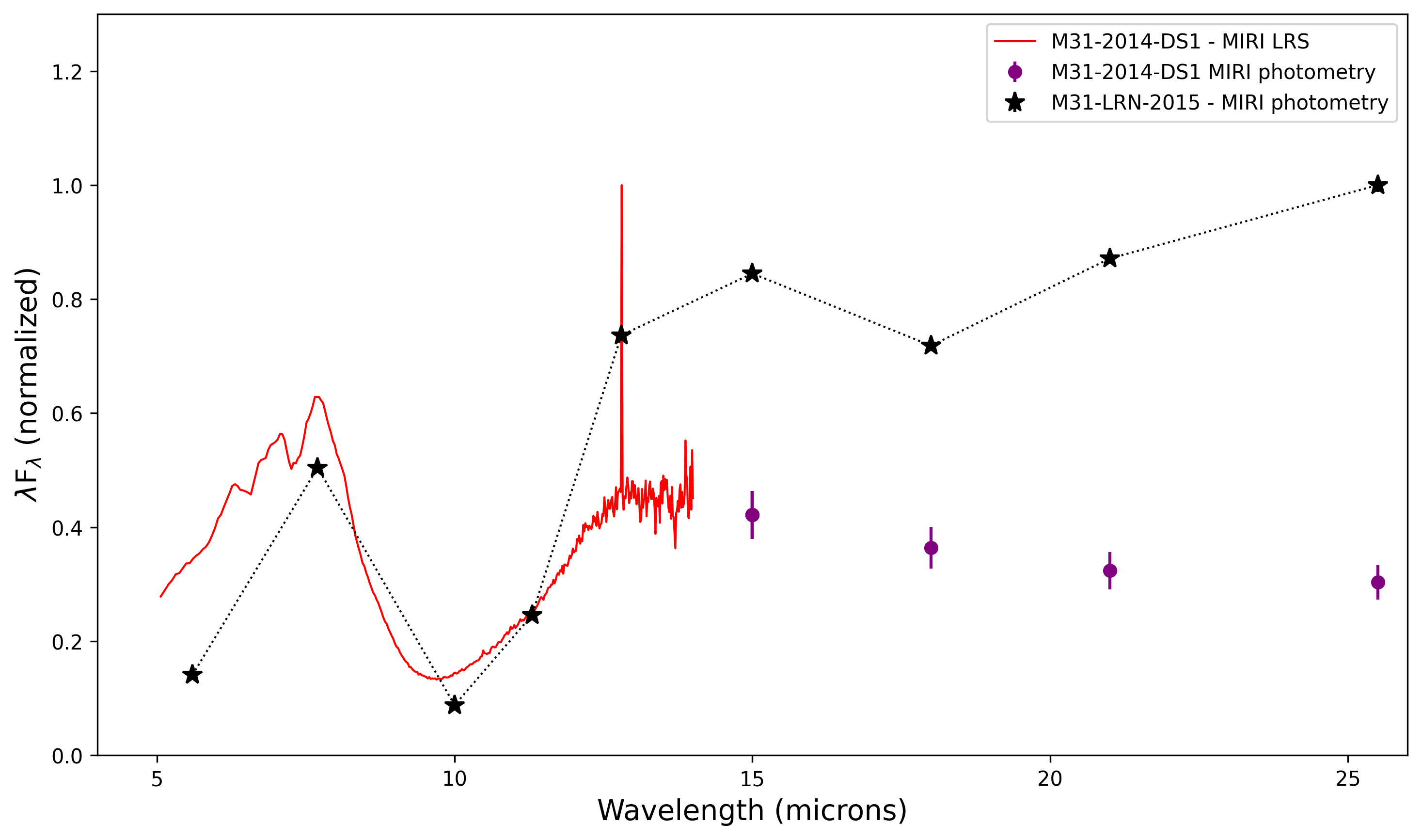}
    \caption{Comparison between the mid-IR appearance of M31-2014-DS1 and luminous red nova M31-LRN-2015 \citep{karambelkar2025hot}. The fluxes have been normalised for direct comparison.}
    \label{fig:lrncomp}
\end{figure}

The mid-infrared spectrum exhibits a strong silicate absorption feature at 10\,$\mu$m, indicative of a substantial amount of cool oxygen-rich circumstellar dust in absorption along the line of sight. In typical red supergiants undergoing standard mass-loss, this feature generally appears in emission \citep[e.g.][]{beasor2020new}. However, in cases with large quantities of circumstellar material---particularly when the dust distribution is asymmetric---the silicate feature can instead appear in absorption, as observed in objects such as IRAS 01304+6211, an OH/IR object in Cassiopeia \citep{kemper2002} and WOH G64, a supergiant known to have a non-spherical dusty torus \citep{ohnaka2008spatially}. \citet{karambelkar2025hot} present JWST observations of luminous red novae several years after their outbursts, including M31-LRN-2015, claimed to be a merger product that show strong silicate absorption features in its spectra, see Fig. \ref{fig:lrncomp}. Additional absorption features are present at 4.3 and 4.5\,$\mu$m, likely associated with CO and possibly CO$_2$, further supporting the presence of a significant amount of dusty molecular material \citep{wood1989co,cox1988co, cami1997co}. We also present the pre-disappearance spectral energy distribution (SED) of the progenitor star for comparison,  as well as HST data from 2022, see Fig. \ref{fig:progenitor}.

When comparing the photometry of the progenitor system to the new JWST data there are clear differences, see Fig. \ref{fig:progenitor}. Firstly, the near-IR flux ($\lambda$ $<$ 2$\mu$m) has faded by a factor of $\sim$5000 at 1$\mu$m and by a factor of 50 since 2022. At longer wavelengths, there is still a loss of flux compared to the 2012 emission, but at a much lower factor. At $\lambda$ $\geq$ 10$\mu$m, the flux has remained constant. The unchanged flux at $\lambda > $ 10$\mu$m is consistent with the system increasing in extinction (likely due to dust production), which dramatically reduces the flux at shorter wavelength and causes the silicate feature to go into absorption rather than emission. The longer wavelength observations are not consistent with an overall reduction in bolometric luminosity of the system. 

If the circumstellar dust surrounding DS1 is asymmetric (see below) then the integrated IR luminosity is only a lower limit to the source's true luminosity, since this method relies on the assumption that any radiation lost to extinction at shorter wavelengths is re-emitted at longer wavelengths. This assumption only holds true for stars with spherically symmetric dust shells. As such, any luminosity estimate derived from such modeling would be a lower limit, as optically thick, asymmetric dust configurations could obscure significant amounts of flux along certain lines of sight, whereas the light being re-emitted at longer wavelengths may be only a small fraction of the total intrinsic luminosity. This hypothesis requires that an Earth-based observer is viewing the system from a vantage point at latitudes near the equatorial plane.

\subsection{DUSTY modeling}
We explore models for the source using the 1-D {\tt DUSTY} radiative transfer code, which solves the radiative transfer equations assuming spherical symmetry. For the dust shell models used in this work, we used MARCS model atmosphere models \citep{gustafsson2008grid} at $T_{\rm eff}$ = 4400 to match the inferred temperature from \citet{de2024ds1}. For the dust itself we use the silicate-rich composition defined by \citet{draine1984optical} and vary the grain size from 0.1 to 1.0$\mu$m in steps of 0.1. We allow the inner dust temperature (i.e. the temperature at the innermost radius of the dust shell) to vary between 100K to 1200K in steps of 100K. We chose 1200K as a maximum $T_{\rm in}$ value since silicate dust sublimates at temperatures above this \citep{hanner1988}. 

For fitting the models, we convolve the model spectrum from {\tt DUSTY} with the JWST filter profiles for the MIRI photometry. We also convolve the NIRSpec and MIRI LRS spectrum with filter profiles for NIRCAM/F360M, MIRI/F560W, and MIRI/F1000W to create synthetic photometry and improve the fit at shorter wavelengths. 

\begin{figure*}
    \centering
    \includegraphics[width=0.8\linewidth]{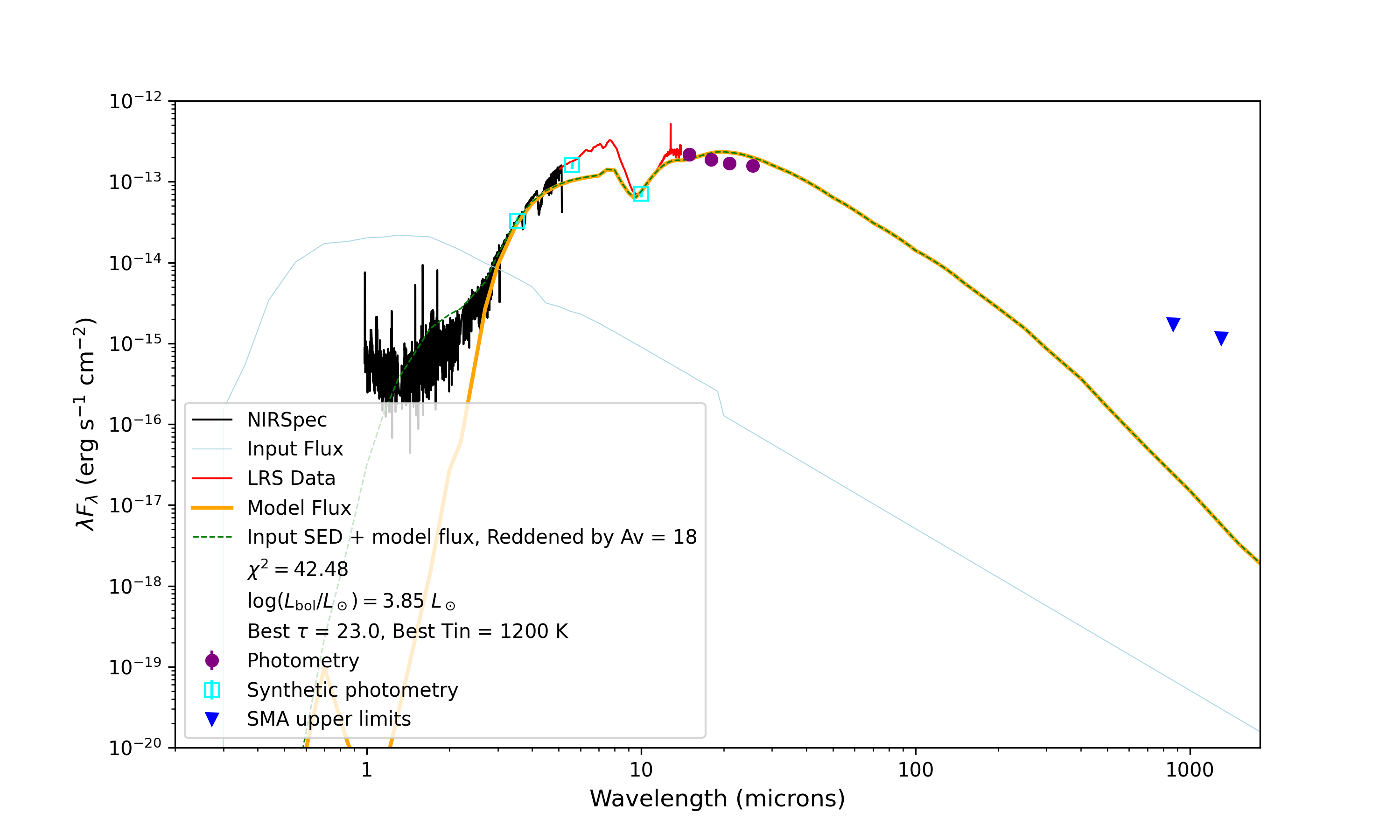}
    \caption{Fit to the JWST data with {\tt DUSTY} modeling and an additional reddened component.}
    \label{fig:dustyfit}
\end{figure*}

We found that no single {\tt DUSTY} model could provide a good fit to the full SED of M31-2014-DS1, see Fig. \ref{fig:dustyfit}. When attempting this, it is possible to fit the SED well at $\lambda$$>$2$\mu$m, but at shorter wavelengths there appears to be a hotter, separate component. As noted above the strong silicate absorption feature and the presence of CO and CO2 features strongly suggest the dust is not spherically symmetric, and so it is expected that using a spherically symmetric radiative transfer code such as {\tt DUSTY} will fail to fit the JWST data. We instead model the SED using two components to mimic the effect that an asymmetric dust distribution would have on the output SED. We begin by fitting only the cooler component of the SED with {\tt DUSTY} as described above. To model the NIRSpec data ($\lambda <$ 2$\mu$m) we take the input model SED ($T_{\rm eff}$ = 5000K, blue line) and redden it by 20 magnitudes using the G23 extinction law \citep{G23extlaw}. We then linearly combine this model component with a scaled-down version of the input spectrum to create a compsite model of the full SED (green dashed line). 

We show our best-fit model in Figure \ref{fig:dustyfit}. The emergent luminosity of the model, after applying 18 mag of extinction, is $\log(L/L_\odot)=3.85$, and we again note that due to the likely asymmetric nature of the dust this is a lower limit on the true luminosity of the system. We find that a high optical depth, $\tau_{V}$ = 23.2 is required to fit the silicate feature in absorption, corresponding to a total dust mass of 0.004\msun. There may be cold dust shells at larger radii that do not effect the near and mid-IR SEDs. 

\section{Discussion}

\subsection{Failed supernova models}

The idea that a massive star could collapse into a black hole without producing a successful supernova explosion was first proposed by \citet{nadezhin1980some}, who suggested that the sudden loss of gravitational binding energy due to neutrino emission during core collapse could unbind the loosely bound hydrogen envelope of a red supergiant. This early insight laid the foundation for later theoretical work by \citet{woosley1986evolution} and \citet{fryer1999mass}, who modeled the outcomes of core collapse in massive stars. While these studies often focused on successful explosions or gamma-ray bursts (GRBs) via the collapsar mechanism, they noted that in cases lacking sufficient rotation or involving extended stellar envelopes, the collapse could fail to launch a strong explosion, resulting instead in a direct black hole formation with minimal electromagnetic emission.

Building on \citet{nadezhin1980some}, \citet{lovegrove2013very} modeled the low-energy transients expected when an RSGs envelope is ejected purely by neutrino-driven mass loss, predicting a faint, short-lived optical signal. More recent work by \citet{o2011black} and \citet{oconnor2011bh} introduced the ``compactness parameter'' as a diagnostic for identifying which stellar progenitors are most likely to fail, providing a quantitative framework for predicting black hole formation in core-collapse events  \cite[see also][]{sukhbold2018high}.

The concept of failed SNe gained further relevance with the emergence of the ``red supergiant problem,'' which noted a lack of high-mass progenitors ($\gtrsim$18--20\,\msun) in pre-explosion imaging \citep{smartt2009death, smartt2015observational}. To test the failed SN hypothesis observationally, \citet{kochanek2008survey} proposed a long-term monitoring survey of nearby galaxies to search for massive stars that disappear without a bright explosion. But the expected observational signature of a failed SN is unclear. Below, we discuss various models that make predictions for the observable features of failed SNe, as well as the bespoke model proposed to explain the observations of M31-2014-DS1. 

\subsubsection{Lovegrove \& Woosley 2013}
In this work, the authors used hydrodynamical models of 15 and 25\msun\ progenitors with weakly bound envelopes (i.e. RSGs) to explore the fates of SN progenitor stars following the sudden loss  of 0.2 - 0.5\msun\ of gravitational mass from their centres via neutrinos. They find that this sudden loss of mass can generate a very weak outgoing shock. For an RSG progenitor, this weak shock is able to eject the hydrogen envelope and form a very weak optical transient with a total kinetic energy of $\sim$10$^{47}$ erg and an ejection velocity of only 100km/s. The models suggest such an event would result in faint, red transients that could last up to 1 year, peaking at a luminosity of approximately 10$^{39}$ erg/s. In-fact, these events would appear similarly to luminous red novae (LRN), though with lower ejection speeds. The observations of M31-2014-DS1 show a lack of any bright optical outburst as predicted by this model. 

\subsubsection{Perna et al. 2013}
While not a study of failed supernovae, in Perna et al. the appearance of compact objects surrounded by disks was explored. They find that for SN that leave behind BHs, the presence of disks is ubiquitous \cite[][]{perna2014disk}. For weak explosions (the case most similar to a failed SN), the outer envelope may remain bound and allow for the formation of a disk. 

The expected duration of associated X-ray emission varies significantly depending on the properties of the fallback disk. For compact, hyper-accreting disks—such as those forming from rapidly rotating, low-metallicity stars typically associated with long-duration GRBs—the transient emission (including X-rays) is expected to last no more than several hundred seconds. In contrast, if the fallback material forms a longer-lived disk, the emission may persist for much longer. This scenario could arise from more slowly rotating but highly extended progenitors, such as supergiant stars, where the large stellar radius compensates for the lower rotational velocity by contributing sufficient angular momentum to sustain an extended accretion disk. or through weak/anisotropic explosions that allow outer layers to remain bound, and the resulting phenomena can have much longer durations. Early phases of these long-lived disks, even if no longer neutrino-cooled, could produce weaker, very long gamma-ray transients, which would likely have associated X-ray emission lasting for several days. For older, long-lived disks around BHs, significant X-ray luminosity, comparable to that of anomalous X-ray pulsars (AXPs) or modeling ultraluminous X-ray sources (ULXs), is expected to persist for much longer periods, potentially on the order of 10$^{4}$ to 10$^{5}$ years. Therefore, the X-ray emission from fallback onto a newly formed black hole can range from brief transient bursts to very long-lasting, persistent sources.

\citet{basinger2021bh1} suggested that observing X-rays may not be a useful diagnostic for failed SNe as the X-rays may be absorbed via the envelope of the collapsed star. For another failed SN candidate, N6946-BH1, no X-rays were detected by Chandra X-ray telescope \citep{basinger2021bh1}. No X-rays were detected at the position of M31-2014-DS1 in archival data \citep{de2024ds1} or more recent observations (this work, see Section \ref{sec:data}). 

\subsubsection{Antoni \& Quataert 2023}
\citet{antoni2023failed} explored the fate of massive stars that fail to produce a successful SN. Specifically, they model a 16.5\msun\ RSG and a lower mass yellow supergiant (YSG) using the {\tt athena++} hydro code. In each case, a large fraction of the stars' envelope would collapse into the core and produce a BH. Importantly, the accretion luminosity from the infall of the envelope would lead to the production of a visible transient. 

In this model, a failed SN would appear similar to a luminous red nova (LRN), with a luminous red plateau lasting hundreds of days. The RSG model produces a plateau length of $\sim$ 500 days, while the YSG plateau lasts 200 days. 

The failed SN candidate N6946-BH1 is also noted as being qualitatively similar to these events, although the models predict a somewhat higher explosion energy and luminosity than observed for N6946-BH1 if its progenitor was a 25\msun\ RSG. The authors also find that an 11.2\msun\ YSG would produce a luminous transient significantly brighter than that of N6946-BH1 (see Fig. 13 within). Given the presumed outburst for M31-2014-DS1 was even less luminous than that of BH1, the observations of M31-2014-DS1 do not fit this model.

\subsubsection{De et al. 2024b}
The observations of M31-2014-DS1 appear to show a scenario that is not fully consistent with any of the aforementioned BH formation models. For one, there is a lack of any bright optical outburst as  predicted by both \citet{lovegrove2013very} and \citet{antoni2023failed}. Secondly, the progenitor itself differs from the typically assumed most-likely failed SNe progenitor. The progenitor to M31-2014-DS1 appeared as a H-depleted YSG progenitor, rather than a H-rich RSG. \citet{de2024ds1} attempt to reconcile the lack of a bright outburst and a failed SN by proposing an intermediate case of a weak impulsive ejection that caused enough additional extinction from new dust formation to counteract the luminosity. They find that the luminosity and duration of any missed outburst is limited to $\leq$ 10$^{5}$\lsun\ and $\leq$ 70 days.  

\citet{lovegrove2013very} attribute the lingering flux from a failed SN to H-recombination in the ejected envelope. In the scenario laid out by \citet{de2024ds1}, there is also H-recombination in addition to flux coming from accretion onto the BH. However, the progenitor system was identified as a hotter, H-depleted star (based on the progenitors' position on the HR diagram), meaning a lower-mass H envelope than studied in the \citet{lovegrove2013very} scenario. In addition, there is a 10$^{48}$ erg shock into the envelope which unbinds only 0.1\msun\ of material. The authors suggest this weak shock would produce a brief, luminous optical transient that would be easy to miss given the photometric coverage of the source. The authors also suggest that convective turbulence during the final stages of evolution lead to an angular momentum that can suppress direct accretion.Combined these effects lead to less efficient accretion (only 1\% of infalling material is likely to accrete directly) and a longer fading time.

The model also addresses the lack of detected X-ray emission from the accreting black hole. No X-rays were found in archival Chandra, Swift, or NuSTAR observations taken between 2015 ({\it Chandra, NuStar}) and 2020 ({\it Swift}). \citet{de2024ds1} explain this by proposing that the newly formed black hole is heavily enshrouded by both the unbound ejecta and a dense outflow from the inefficient accretion process. This material creates a high column density that absorbs any X-rays, preventing their detection. The authors predict that this obscuring material will dissipate over time, and the source may become transparent to soft X-rays within the next several years to decades, potentially revealing the nascent black hole through future observations. At the time of the most recent {\it Chandra} observations, taken in November 2024, the source has still not been detected, see Section \ref{sec:data}.

\section{Is M31-2014-DS1 a failed supernovae?}
 In the previous sections we discussed the predicted observational signatures for a failed SN from a number of models. Ultimately, the only model that appears to fit the data is that of \citet{de2024ds1}. This bespoke model was constructed specifically to reconcile the observations of M31-2014-DS1 with the failed supernova scenario, proposing that a weak shock in a hydrogen-depleted progenitor star could explain the lack of a bright optical outburst.

However, several observational details challenge the interpretation of M31-2014-DS1 as a failed SN. Ten years after the initial mid-infrared brightening in 2014, JWST observations reveal a persistent, luminous infrared source at its location. The long-wavelength flux ($\lambda \geq 10\,\mu\text{m}$) has remained nearly constant since 2012, which is inconsistent with a simple, overall reduction in the system's bolometric luminosity. In a failed SN scenario where a BH is formed, the luminosity from fallback accretion is expected to decline over time as the accretion rate decreases. The fact that the source's mid-IR flux has not faded significantly, and that its current luminosity is estimated to be around 7--25\% of the progenitor's, complicates the collapse model.

Furthermore, no X-rays have been detected from the position of M31-2014-DS1 in archival Chandra observations from 2015, Swift observations from 2020 or Chandra observations from 2024. This is in contention with models that predict long-lasting X-ray emission from fallback accretion onto a new black hole \citep[e.g.][]{perna2014disk}, potentially for thousands of years. Although the non-detection could be explained by heavy obscuration from the collapsing star's envelope or newly formed dust \citep{basinger2021bh1,de2024ds1}, the combination of a lingering, steady infrared flux and a lack of detectable X-rays more than 10 years after the suggested collapse of the star would argue against a failed SN.

\section{Is M31-2014-DS1 a stellar merger?}
As demonstrated in the previous section, M31-2014-DS1 does not fit any of the standard predictions for a failed SN. The only model that successfully matches the observations is the custom model constructed for this object from \citet{de2024ds1}. While the authors do make a prediction for the fate of the source, i.e. that it will continue to fade below it's pre-disappearance luminosity, this alone is not sufficient in confirming the failed SN scenario.  A difficulty with such revised failed SN models is that by invoking long-lasting accretion to explain the lingering late-time luminosity, one pushes the observed expectations to a regime that becomes difficult to distinguish from a range non-terminal outburst scenarios, such as dusty shell ejection and stellar merger events.  There have been a number of observed eruptive transients that have faded in the optical but remained bright in the infrared as they became dust obscured, many of which are interpreted as stellar merger events (e.g.\: V1309~Sco; \citep{Tylenda2011}, V4332~Sgr; \citep{Kaminski2010}, M85 OT2006‑1; \citep{Rau2007}, M31 RV; \citep{Bond2011}, NGC 4490‑OT2011; \citep{Smith2016}, M101 OT2015‑1; \citep{Blagorodnova2017}, SPIRITS IR‑transient sample; \citep{Jencson2019}, JWST‑studied mergers (AT2021blu, AT2021biy, AT2018bwo, M31‑LRN‑2015); \citep{karambelkar2025hot}).

In the binary merger scenario, the observed luminosity of any progenitor system can have a significant contribution from shock heating of the outflow from the inspiral phase, raising the apparent luminosity \citep{pejcha2017,smith2018merger}. In other words, the observed yellow supergiant progenitor luminosity is not the nuclear burning luminosity of the system, but is enhanced by the inspiral phase. This luminosity is not indicative of an initial mass using the usual mass-luminosity relations from single star models; the true stellar luminosity and corresponding initial mass are lower. As such, converting from an observed luminosity to an initial mass would lead to an overestimate of the progenitor mass \citep[see][]{beasor2024BH1}. In addition, this means that the apparent drop in luminosity observed for both M31-2014-DS1 and N6946-BH1 is not necessarily entirely due to extinction from an ejected envelope or dust production, since the luminosity may naturally decrease as the extra luminosity from the inspiral phase is shut off -- the system may actually be returning to it's quiescent radiative luminosity. This pre-merger increase in luminosity due to inspiral was observed for the merger system V1309 Sco \citep{Tylenda2011}. 

\citet{kashisoker17} propose that stellar mergers, particularly those classified as Type II intermediate-luminosity optical transients (ILOTs), can exhibit such behavior due to anisotropic mass loss. In their model, a strongly interacting binary ejects mass preferentially into the equatorial plane, forming a dense, optically thick dust torus that obscures the central source from equatorial viewing angles. Observers in these directions predominantly detect reprocessed radiation from the polar regions, leading to a dramatic reduction in optical flux (down to $\sim$10\% of its original value), while the infrared emission remains strong due to dust heating. It is important to recognize a clear selection effect that favors the identification of edge-on systems in surveys searching for disappearing stars. In such merger events, the observed fading of the progenitor is typically caused by obscuration from an equatorial dust torus. Consequently, merging systems are more likely to be misidentified as failed supernova candidates when they are viewed nearly edge-on, because systems that are nearly edge-on will fade the most dramatically. Face-on mergers are expected to occur as well, but they are less likely to exhibit significant fading and therefore are less likely to be flagged as a disappearing star \citep{andrews2021}. The intrinsic luminosity of an ILOT outburst likely depends on the mass of the engulfed companion, and in edge-on systems, extinction could obscure a significant portion of the outburst’s light—especially if the event was relatively weak.

\citeauthor{kashisoker17} outline five evolutionary phases for such events: (1) \textit{Pre-outburst interaction}, involving gradual brightening and increasing IR-to-optical luminosity ratios as dust begins to form; (2) \textit{Early outburst}, characterized by a sharp rise in both optical and IR luminosity as mass is ejected; (3) \textit{Late outburst}, during which equatorial dust fully obscures the central source, causing significant optical dimming; (4) \textit{Post-outburst}, where the system is dominated by IR emission from polar dust; and (5) \textit{Recovery}, as the dust shell gradually thins and the central source becomes visible again over years to decades.

The predicted observational signatures of Type II ILOTs can closely resemble those of failed supernovae, particularly the apparent disappearance of a massive star in visible wavelengths. Both M31-2014-DS1 and NGC~6946-BH1, for example, exhibit this behavior. In the ILOT scenario, this fading is explained by asymmetric dust formation obscuring the source, rather than by collapse to a black hole. Notably, some failed supernova candidates, such as NGC~6946-BH1, also show a luminous optical outburst ($\sim$$10^6$~$L_\odot$) prior to their disappearance, consistent with the ``Early outburst'' phase of the ILOT model. The extended duration of the optical dimming—lasting decades—is likewise consistent with the long recovery timescales predicted for equatorially enshrouded merger remnants. Kashi \& Soker specifically suggest that NGC~6946-BH1 could be a Type II ILOT and predict that its visible emission may eventually re-emerge as the dust clears over time.

We note however, that predictions for the appearance of mergers and merger products are generally lacking through the literature. The presence of any outburst may depend on the mass of the companion as well as the orientation of the system. Qualitatively, the behaviour of both M31-2014-DS1 and N6946-BH1 seem to be consistent with observations of mergers (e.g. V1309 Sco). For DS1, we note that it's JWST spectrum shares a number of features with observations of luminous red novae \citep{karambelkar2025hot}, in particular M31-LRN-2015, which is thought to have been the result of a merger.

\section{Conclusions}

In this paper, we investigate the likely fate of the failed supernova candidate M31-2014-DS1 in light of new observations from JWST and the SMA. Our main findings are as follows:

\begin{itemize}
\item We detect an infrared source coincident with the position of M31-2014-DS1 in JWST/NIRSpec, MIRI/LRS, and MIRI imaging. We derive a luminosity of $\log(L/L_\odot) = 3.85$, corresponding to approximately 7\% of the progenitor system identified by \citet{de2024ds1}. This value is likely a lower limit, as the spectral energy distribution (SED) shows the 10~$\mu$m silicate feature in absorption, alongside other molecular absorption features, indicative of high extinction along our viewing direction and an asymmetric dust geometry. We also cannot know whether the luminosity of the source identified by \citet{de2024ds1} was the quiescent luminosity of the system, or whether this was a value heightened by binary processes such as inspiralling. 
\item No source is detected at radio wavelengths.
\item No X-rays are detected. 
\item We compare the observed properties to theoretical predictions for the outcome of a failed supernova. Current models, however, provide no clear expectations that would allow a failed supernova remnant to be distinguished observationally from a stellar merger.

\end{itemize}

Ultimately, while the source is significantly fainter than the progenitor, its fate remains uncertain. One possibility is that the star has collapsed to a black hole, in which case it will continue to fade. Alternatively, the event may have been the result of a stellar merger, in which case the system could re-brighten once the dust obscuration diminishes and the central source becomes visible. Distinguishing between these scenarios will require further JWST monitoring over the coming years.

\section*{Acknowledgements}

E.R.B. is supported by a Royal Society Dorothy Hodgkin Fellowship (grant no. DHF-R1-241114). Time-domain research by the University of Arizona team and D.J.S. is supported by National Science Foundation (NSF) grants 2108032, 2308181, 2407566, and 2432036 and the Heising-Simons Foundation under grant \#2020-1864. 
\section*{Data Availability}

All JWST data presented in this work are available via MAST. The SMA data are available by request. Chandra data are available from the Chandra Data Archive (CDA).



\bibliographystyle{mnras}
\bibliography{references} 





\bsp	
\label{lastpage}
\end{document}